\documentstyle[preprint,aps]{revtex}
\tightenlines
\input epsf.tex
\def\DESepsf(#1 width #2){\epsfxsize=#2 \epsfbox{#1}}
\begin{document}
\preprint{\vbox{\hbox{COLO-HEP-434}\hbox{hep-ph/9908531}}} \draft
\title{Combined study of $b \rightarrow s \gamma$ and the muon anomalous magnetic moment in gauge-mediated
supersymmetry breaking models}
\author{K. T. Mahanthappa$^1$\footnote{ktm@verb.colorado.edu} and Sechul Oh$^2$\footnote{scoh@kimcs.yonsei.ac.kr} }
\address{$^1$ Department of Physics, University of Colorado, Boulder, Colorado 80309, USA  \\
$^2$ Institute of Physics and Applied Physics, Yonsei University, Seoul, 120-749, Korea}
\date{August, 1999}
\maketitle
\begin{abstract}
We study both the branching ratio for $b \rightarrow s \gamma$ decay and the muon anomalous magnetic moment,
$a_{\mu} \equiv (g-2)_{\mu} /2$, in the minimal supersymmetric standard model with gauge-mediated supersymmetry
breaking. Combining new experimental data on $a_{\mu}$ and the branching ratio for $b \rightarrow s \gamma$,
strong limits on the parameter space of these models are derived. We find that this combined study leads to much
stronger constraints on the parameter space of the model than those from either $b \rightarrow s \gamma$ or
$a_{\mu}$.  In particular, the region of large $\tan \beta$ is extremely limited, which would have been otherwise
allowed. We include the supersymmetric one-loop correction to the mass of $b$ quark, $m_b$, and find that in order
to have a correct value of $m_b$, the region of large tan$\beta$ and $\mu <0$ (in our convention) is not allowed
in these models. The region of large tan$\beta$ and $\mu >0$ is also strongly constrained. We present bounds on
supersymmetric particle masses as a function of $\tan \beta$.
\end{abstract}

\newpage
Supersymmetry (SUSY) is a very attractive candidate beyond the standard model (SM) since it provides an elegant
solution to the hierarchy problem in particle physics.  Understanding the mechanism of SUSY breaking and its
communication to the observable sector is one of the most important open questions.  There are two types of
realistic supersymmetric models of interest : gravity-mediated models and models with gauge-mediated SUSY breaking
(GMSB).  The GMSB models \cite{1} have been of special interest, because they have attractive features of natural
suppression of the SUSY contributions to flavor-changing neutral currents at low energies and prediction of the
supersymmetric particle mass spectrum in terms of few parameters.

The parameters of SUSY models can be constrained by using precision measurements in low energy experiments,
because superparticles contribute to low energy physics through radiative corrections.  In particular,
experimentally observed rare decays may shed light on the parameter space of SUSY models.  Processes such as $b
\rightarrow s \gamma$ do not occur at the tree level, and at one-loop level they occur at a small rate but enough
to be sensitive to new physics effects \cite{bsgamma}. The CLEO collaboration has recently reported the branching
ratio for the decay $b \rightarrow s \gamma$ \cite{2} : ${\rm BR} (b \rightarrow s \gamma) = (3.15 \pm 0.35 \pm
0.32 \pm 0.26) \times 10^{-4}$, which corresponds to the bound $2.0 \times 10^{-4} < {\rm BR} (b \rightarrow
s\gamma) < 4.5 \times 10^{-4}$ at 95 $\%$ C.L.  The anomalous magnetic moment of muon, $a_{\mu} \equiv (g-2)_{\mu}
/2$, is also sensitive to new physics effects and can be used to constrain SUSY models \cite{7,7a,7b}, on account
of the great accuracy of both experimental and SM theoretical values of $a_{\mu}$.  The present experimental value
of $a_{\mu}$ \cite{3} is $a^{\rm exp}_{\mu} = 11659230(84) \times 10^{-10}$, while the theoretical prediction for
$a_{\mu}$ in  the context of the SM \cite{4,5} is $a^{\rm SM}_{\mu} = 11659162(6.5) \times 10^{-10}$.

In this paper, we obtain combined constraints due to both $b \rightarrow s \gamma$ decay and $a_{\mu}$ in the
minimal supersymmetric SM (MSSM) with GMSB.  Even though there exist the previous works which studied either $b
\rightarrow s \gamma$ \cite{6,6a} or $a_{\mu}$ \cite{7b} in the GMSB models, our work extends the previous ones in
the sense that we investigate both $b \rightarrow s \gamma$ and $a_{\mu}$ together with the inclusion of the
supersymmetric one-loop correction to the mass of $b$ quark, $m_b$, which has considerable effects in large $\tan
\beta$ region \cite{8,8a}.  We shall see that this combined study leads to much stronger constraints on the
parameter space of the model than those from either $b \rightarrow s \gamma$ or $a_{\mu}$.  In particular, the
region of large $\tan \beta$ is extremely limited, which would have been otherwise allowed. Furthermore, in this
work, we explicitly show that, with the presently available experimental data, constraints from the decay $b
\rightarrow s \gamma$ are more stringent than those from $a_{\mu}$ in broad region of the parameter space. We also
present bounds on supersymmetric particle masses as a function of $\tan \beta$.

In the GMSB models messenger fields transmit SUSY breaking to the fields of visible sector via loop diagrams
involving SU(3)$_C \times$SU(2)$_L \times$U(1)$_Y$ gauge interactions.  The simplest model consists of messenger
fields which transform as a single flavor of vectorlike ${\bf 5} +\bar {\bf 5}$ of SU(5).  These messenger fields
may be coupled to a SM singlet chiral superfield $S$ through the superpotential
\begin{eqnarray}
W_{messenger} = \lambda_D S D \bar D + \lambda_L S L \bar L,
\end{eqnarray}
where the fields have the SM representations and quantum numbers $D:({\bf 3},{\bf 1})_{Y=-2/3}$, $\bar D:(\bar
{\bf 3},{\bf 1})_{Y=2/3}$, $L:({\bf 1},{\bf 2})_{Y=-1}$, and $\bar L:({\bf 1},{\bf 2})_{Y=1}$. The scalar and $F$
components of $S$ acquire VEVs $\left< S \right>$ and $\left< F_S \right>$, respectively, through their
interactions with the fields of hidden sector, which results in breakdown of SUSY. It is known that for messenger
fields in complete SU(5) representation, at most four (${\bf 5} +\bar {\bf 5}$) pairs, or one (${\bf 5} +\bar {\bf
5}$) and one (${\bf 10} +\overline{{\bf 10}}$) pair are allowed to ensure that the gauge couplings remain
perturbative up to the grand unified theory (GUT) scale \cite{9}.

In general, the parameters $\mu$ and $B$ in soft SUSY breaking terms depend on the details of the SUSY breaking in
the hidden sector. We require that electroweak symmetry be radiatively broken, which determines $\mu^2$ and $B$ in
terms of other parameters of the theory. Then the sparticle masses depend on five independent parameters: $M$,
$\Lambda$, $n$, $\tan \beta$, and sign$(\mu)$.  Here $M$ is the messenger scale given by $M = \lambda \left< S
\right>$ with a universal Yukawa coupling $\lambda$ in the messenger sector at GUT scale.  The parameter $\Lambda$
is defined by $\Lambda = \left< F_S \right> / \left< S \right>$. The integer number $n$ is the effective number of
messenger fields given by $n =n_5 +3 n_{10}$, where $n_5$ and $n_{10}$ denote the number of (${\bf 5} +\bar {\bf
5}$) and (${\bf 10} +\overline{{\bf 10}}$) pairs, respectively. The radiatively generated soft SUSY-breaking
masses of gaugino and scalars, $\tilde M_i$ and $\tilde m^2$, at messenger scale $M$ are given by \cite{10,10a}
\begin{eqnarray}
\tilde M_i(M) = n g(x) {\alpha_i(M)\over4\pi} \, \Lambda , \label{Mi}
\end{eqnarray}
\begin{eqnarray}
\tilde m^2(M) = 2 n f(x) \sum_{i=1}^3 k_i C_i \biggl({\alpha_i(M)\over4\pi}\biggr)^2 \Lambda^2 , \label{m2}
\end{eqnarray}
where $x \equiv \Lambda / M$. $\alpha_i$ $(i=1,2,3)$ are the three SM gauge couplings with GUT normalization for
$\alpha_1$.  $k_i$ are 1, 1, 3/5 for SU(3)$_C$, SU(2)$_L$ and U(1)$_Y$, respectively.  $C_i$ are zero for gauge
singlets, and 4/3, 3/4 and $(Y/2)^2$ for the fundamental representations of SU(3)$_C$, SU(2)$_L$ and U(1)$_Y$,
respectively (with $Y$ defined by $Q= I_3 + Y/2$). $g(x)$ and $f(x)$ are messenger scale threshold functions.

We use the input values $\alpha_s(M_Z)=0.118$, $\sin^2 \theta_W (M_Z)=0.2315$ and $\alpha(M_Z)=1/128$. The
parameter $\Lambda$ is taken to be around 100 TeV to ensure that the sparticle masses are of the order of the weak
scale. We restrict $1 <M / \Lambda < 10^4$ : the case $M = \Lambda$ is excluded since it produces a massless
scalar in the messenger sector \cite{10,10a} and the upper bound on the gravitino mass of about $10^4$ eV
restricts $M / \Lambda < 10^4$ \cite{10a,11}. Using the appropriate renormalization group equations (RGEs)
\cite{12}, we first go up to the messenger scale $M$ with gauge and Yukawa couplings, and fix the sparticle masses
with the boundary conditions (\ref{Mi}) and (\ref{m2}).  We next go down with the $6 \times 6$ mass matrices for
the squarks and sleptons to find the sparticle spectrum. In running the RGEs, we include the one-loop correction
to the running bottom quark mass, $\Delta m_b$, which involves the contributions coming from
gluino$-$bottom-squark loop diagram and chargino$-$top-squark loop diagram, and is given by \cite{8}
\begin{eqnarray}
\Delta m_b = - \lambda_b v_1 \mu \tan\beta \left[ {2\alpha_s \over 3\pi} M_{\tilde g} I(m^2_{\tilde b_1},
m^2_{\tilde b_2}, M^2_{\tilde g}) + {\lambda^2_t \over (4\pi)^2} A_t I(m^2_{\tilde t_1}, m^2_{\tilde t_2}, \mu^2)
\right], \label{mb}
\end{eqnarray}
where the integral function $I(a,b,c)$ is given by
\begin{eqnarray}
I(a,b,c) = -{ab \ln(a/b) +bc \ln(b/c) +ca \ln(c/a) \over (a-b)(b-c)(c-a)},
\end{eqnarray}
and $M_{\tilde g}$ is the gluino mass and $m_{\tilde b_i}$ ($m_{\tilde t_i}$) is  the bottom-squark (top-squark)
eigenstate masses, respectively. Our convention for the sign of the Higgs boson mass parameter $\mu$ follows that
in Ref. \cite{12}.

Calculation of $b \rightarrow s\gamma$ amplitude involves the coefficients of short distance photonic and gluonic
operators $c_7(M_W)$ and $c_8(M_W)$. The SM calculations of $c_7$ and $c_8$ with QCD corrections to two-loop order
are given in Ref. \cite{13}.  Calculations of next-to-leading order agree with the previous calculations while
reducing the theoretical errors \cite{14}.   Various supersymmetric contributions for $b \rightarrow s\gamma$ are
given in a generic form in Ref.\cite{15}. The branching ratio for $b \rightarrow s \gamma$ is given by \cite{16}
\begin{eqnarray}
{\rm BR} (b \rightarrow s \gamma) = {6\alpha \over \pi} {[\eta^{16/23} A_{\gamma} +{8 \over 3} (\eta^{14/23}
-\eta^{16/23}) A_g + C]^2 \over I(m_c /m_b) [1 -{2 \over 3\pi} \alpha_s (m_b) f(m_c /m_b)] } {\rm BR} (b
\rightarrow ce \bar \nu) ,
\end{eqnarray}
where $\eta = \alpha_s (M_Z) / \alpha_s (m_b)$.  $I$ is the phase-space factor given by $I(x) = 1 -8 x^2 +8 x^6
-x^8 -24 x^4 \ln x$, and $f$ is the QCD correction factor for the semileptonic decay given by $f(m_c / m_b) =
2.41$.  $A_{\gamma}$ and $A_g$ are the coefficients of the effective $bs\gamma$ and $bsg$ penguin operators
evaluated at the scale $M_Z$, respectively. ${\rm BR} (b \rightarrow ce \bar \nu)$ denote the branching ratio of
the semileptonic decay $b \rightarrow ce \bar \nu$.

The supersymmetric contributions to the muon anomalous magnetic moment $a_{\mu}$ are essentially coming from
neutralino($\tilde \chi^0$)-smuon($\tilde \mu$) loop diagram and chargino($\tilde \chi^+$)-sneutrino($\tilde \nu$)
loop diagram as follows \cite{7a,7b} :
\begin{eqnarray}
\delta a^{\rm SUSY}_{\mu} = \delta a^{\rm N}_{\mu} + \delta a^{\rm C}_{\mu} .
\end{eqnarray}
Here $\delta a^{\rm N}_{\mu}$ and $\delta a^{\rm C}_{\mu}$ denote the contributions from the $\tilde
\chi^0$-$\tilde \mu$ diagram and $\tilde \chi^+$-$\tilde \nu$ diagram, respectively, and are given by
\begin{eqnarray}
\delta a^{\rm N}_{\mu} &=& {m_{\mu} \over 16\pi^2} \sum_{i,\alpha} \left[ -{m_{\mu} \over 6 m^2_{\tilde \mu_i} (1
-x_{i\alpha})^4} (N^L_{i\alpha} N^L_{i\alpha} + N^R_{i\alpha} N^R_{i\alpha}) (1 -6 x_{i\alpha} +3 x^2_{i\alpha} +2
x^3_{i\alpha} -6 x^2_{i\alpha} \ln x_{i\alpha})   \right. \nonumber \\ &-& \left. {m_{\tilde \chi^0_\alpha} \over
m^2_{\tilde \mu_i} (1 -x_{i\alpha})^3} N^L_{i\alpha} N^R_{i\alpha} (1 -x^2_{i\alpha} +2 x_{i\alpha} \ln
x_{i\alpha}) \right], \\ \delta a^{\rm C}_{\mu} &=& {m_{\mu} \over 16\pi^2} \sum_{l} \left[ {m_{\mu} \over 3
m^2_{\tilde \nu} (1 -x_l)^4}  (C^L_l C^L_l + C^R_l C^R_l) \left( 1 -{3 \over 2} x_l -3 x^2_l +{1 \over 2} x^3_l +3
x_l \ln x_l \right) \right.  \nonumber \\ &-& \left. {3 m_{\tilde \chi^+_l} \over m^2_{\tilde \nu} (1 -x_l)^3}
C^L_l C^R_l \left( 1 - {4 \over 3} x_l + {1 \over 3} x^2_l +{2 \over 3} \ln x_l \right) \right],
\end{eqnarray}
where  ($i, l =1,2$; $\alpha =1 - 4$)
\begin{eqnarray}
x_{i\alpha} &=& m^2_{\tilde \chi^0_{\alpha}} / m^2_{\tilde \mu_i},  \,\,\, x_l = m^2_{\tilde \chi^+_l} /
m^2_{\tilde \nu},  \nonumber \\ N^L_{i\alpha} &=& - {m_{\mu} \over v_1} (U^N)_{3\alpha} (U^{\tilde \mu})_{Li} +
\sqrt{2} g_Y (U^N)_{1\alpha} (U^{\tilde \mu})_{Ri},  \nonumber \\ N^R_{i\alpha} &=& - {m_{\mu} \over v_1}
(U^N)_{3\alpha} (U^{\tilde \mu})_{Ri} - {g_2 \over \sqrt{2}} (U^N)_{2\alpha} (U^{\tilde \mu})_{Li} - {g_Y \over
\sqrt{2}} (U^N)_{1\alpha} (U^{\tilde \mu})_{Li},  \nonumber \\ C^L_l &=& {m_{\mu} \over v_1} (U^C)_{l2}, \,\,\,
C^R_L = - g_2 (V^C)_{l1}.
\end{eqnarray}
$g_2$ and $g_Y$ are the gauge couplings of SU(2)$_L$ and U(1)$_Y$, respectively, and $v_1$ is the vacuum
expectation value of the Higgs boson $H_1$.  $m_{\mu}$, $m_{\tilde \chi^0}$, $m_{\tilde \chi^+}$, $m_{\tilde
\mu}$, and $m_{\tilde \nu}$ are masses of the muon, neutralino, chargino, smuon, and sneutrino, respectively.
$U^N$ and $U^{\tilde \mu}_{L,R}$ denote the neutralino and smuon mixing matrices, and $U^C$ and $V^C$ denote the
chargino mixing matrices.

We use our calculated mass spectrum and couplings to calculate the rate for $b \rightarrow s\gamma$ and $\delta
a^{\rm SUSY}_{\mu}$. For fixed values of $\tan \beta$, $n$ and sgn$(\mu)$, both the branching ratio for $b
\rightarrow s\gamma$ and $\delta a^{\rm SUSY}_{\mu}$ as well as $|\mu|$ and the weak gaugino mass $M_2$ are
calculated, as the values of $M$ and $\Lambda$ vary. Then the bounds on the branching ratio for $b \rightarrow
s\gamma$ and $\delta a^{\rm SUSY}_{\mu}$ are translated into the bounds on values of $M_2$ and $|\mu|$ in the
$|\mu| - M_2$ plane for fixed values of $\tan \beta$, $n$ and sgn$(\mu)$.  From Eq. (\ref{Mi}) one can see that
$M_2$ is directly related to $\Lambda$, since $g(x) \simeq 1$. Bounds on other sparticle masses can be easily
deduced from a bound on $M_2$, owing to the relations Eqs. (\ref{Mi}) and (\ref{m2}).

In $b \rightarrow s\gamma$ decay, the contributions to the total decay amplitude are coming from the $W$ loop
diagram, charged Higgs boson loop diagram, neutralino loop diagram, and gluino loop diagram.  It has been pointed
out that the neutralino and gluino contributions to the amplitude are less than 1 $\%$ in the whole range of
parameter space \cite{6a}.  The charged Higgs boson loop contribution adds constructively to the $W$ loop
contribution, while the chargino loop contribution can be constructive or destructive to the $W$ loop
contribution, but is generally much smaller than the charged Higgs boson loop contribution. We use the new CLEO
bound on the branching ratio for $b \rightarrow s\gamma$ decay in order to obtain constraints on the parameter
space of the GMSB models.

The bound on the supersymmetric contributions to $a_{\mu}$ is given by $-71 \times 10^{-10} < \delta a^{\rm
SUSY}_{\mu} < 207 \times 10^{-10}$ at 90 $\%$ C.L.  This bound is obtained by the difference between experimental
value and theoretical prediction of $a_{\mu}$. The new E821 experiment at Brookhaven is expected to improve the
experimental determination of $a_{\mu}$ to the level of $4 \times 10^{-10}$ \cite{17}. Since the complete two-loop
electroweak contribution in the SM to $a_{\mu}$ is $a^{\rm EW}_{\mu} = 15.1(0.4) \times 10^{-10}$ \cite{5} and the
supersymmetric contributions can be as large or even larger than the $a^{\rm EW}_{\mu}$, it is expected that the
new E821 experiment will be possible to test both the SM electroweak and supersymmetric contributions
\cite{7,7a,7b}.

In Figs. 1$-$8, we display the bounds obtained from the branching ratio for $b \rightarrow s\gamma$ and $\delta
a^{\rm SUSY}_{\mu}$ in the $|\mu| - M_2$ plane for either sign of $\mu$, for $\tan \beta = 10$ and 60, and for
$n=1$ and 3, respectively. Solid lines represent the bounds from the branching ratio for $b \rightarrow s\gamma$
and dot-dashed lines describe the bounds from $\delta a^{\rm SUSY}_{\mu}$. Figures 1 and 2 show the bounds on
$M_2$ and $|\mu|$ for $\tan \beta =10$ and $n=1$, and for positive and negative $\mu$, respectively.  The region
surrounded by the solid line is allowed by the CLEO bound, while the upper region of the dot-dashed line is
allowed by the present bound on  $a_{\mu}$.  In the case of Fig. 1, the constraint from $b \rightarrow s\gamma$
decay is clearly much stronger than that from $a_{\mu}$.  We find $M_2 > 248$ GeV and $\mu > 626$ GeV.  Small
values of $M_2$ lead to unacceptably large contribution to the branching ratio for $b \rightarrow s\gamma$, while
large values of $\mu$ raise the problem of fine-tuning and are generally constrained by the lower bound on the
stau mass. In Fig. 2 we see the constrains from both $b \rightarrow s\gamma$ and $a_{\mu}$ are complementary.  By
combining the bounds from the both, we can obtain much stronger bound on $M_2$ and $|\mu|$; in particular, low
values of $|\mu|$ which would have been allowed are excluded.  We find $M_2 > 210$ GeV and $|\mu| > 505$ GeV.

In large $\tan \beta$ case, we find that the bound from either $b \rightarrow s\gamma$ or $a_{\mu}$ is more
stringent than that in small $\tan \beta$ case, and most region in the $|\mu| - M_2$ plane is excluded.  For $\tan
\beta = 60$ and $\mu >0$ (Figs. 3 and 6), the allowed regions from each of $b \rightarrow s\gamma$ and $a_{\mu}$
do not overlap, even though a possibility exists that they might overlap for unacceptably very large values of
$\mu$. Thus, this case is excluded, while it would be allowed if one considered only either $b \rightarrow
s\gamma$ or $a_{\mu}$ as in Refs. \cite{7b,6a}. For $\tan \beta \lesssim 50$ and $\mu >0$, the allowed regions
from each of $b \rightarrow s\gamma$ and $a_{\mu}$ overlap allowing limited regions in the parameter space. For
$\tan \beta = 60$ and $\mu <0$, the supersymmetric one-loop correction to bottom quark mass leads to unacceptably
large value of $m_b$.  In other words, in order to have a correct value of $m_b$, the negative sign of $\mu$ is
not physically allowed for large $\tan \beta$ in our analysis.  It can be qualitatively understood from Eq.
(\ref{mb}) : for large $\tan \beta$, the correction $\Delta m_b$ is large, and for negative $\mu$, $\Delta m_b$
gives positive contribution leading to an incorrect value of $m_b$. Thus, by inclusion of the correction $\Delta
m_b$, we exclude the case of large $\tan \beta$ and $\mu <0$. This result is different from those in the previous
works : in Ref. \cite{6a} the one-loop correction $\Delta m_b$ was not taken into account and the case of large
$\tan \beta$ and $\mu <0$ was favored without any constraint from $b \rightarrow s \gamma$.  Also, in Ref.
\cite{7b}, the case of large $\tan \beta$ and $\mu <0$ is allowed with a constraint on the value of $M_2$.

For $n=3$, the constraints from each of $b \rightarrow s\gamma$ and $a_{\mu}$ are more stringent.  But after
combining the constraints together, we find that for $\tan \beta =10$ and $\mu <0$, the resulting limits are lower
(i.e., $M_2 > 182$ GeV and $|\mu| > 320$ GeV) than those in the case of $n=1$, while for $\tan \beta =10$ and $\mu
>0$, the resulting limits are still higher (i.e., $M_2 > 357$ GeV and $\mu > 648$ GeV).  For $n=3$, large $\tan
\beta$ region is almost ruled out due to the same reason as the case of $n=1$.

In Figs. 7 and 8 we plot the bounds on the sparticle masses, obtained by this combined analysis of $b \rightarrow
s\gamma$ and $a_{\mu}$, as a function of $\tan \beta$ for positive $\mu$ and for $n=1$ and 3, respectively.  The
plots are displayed for up to $\tan \beta \approx 50$, since the region corresponding to $\tan \beta \gtrsim 50$
is ruled out. The lower bounds on the sparticle masses increase monotonically as $\tan \beta$ does.  For $n=3$,
the lower bound on each sparticle mass is higher than that for $n=1$. (We show the plots in the positive $\mu$
case only, because for negative $\mu$ the allowed parameter space is restricted to only relatively small $\tan
\beta$ region ($\tan \beta \lesssim 20$) in order to have a correct value of $m_b$ in our analysis.)

Some comments concerning the non-leading order (NLO) corrections to $b \rightarrow s \gamma$ are in order.  It has
been pointed out by Kagan and Neubert \cite{23} that NLO corrections lead to a 10$\%$ error in predicted
theoretical values.  Ciuchini {\it et al.} \cite{24} find that, when the leading order (LO) corrections cancel,
contributions from NLO corrections become significant; this happens when the masses of charginos and right-handed
top-squark are very much higher than those of other squarks and the gluino.  In our parametric space such a
cancellation of LO corrections is unlikely.  There could be other circumstances leading to the cancellation of LO
corrections, and this is under investigation.

In conclusion, we have investigated both the $b \rightarrow s\gamma$ decay and the anomalous magnetic moment of
muon $a_{\mu}$ together in MSSM with GMSB.  We have used the new CLEO bound on the branching ratio for $b
\rightarrow s\gamma$ and the present experimental limit on $a_{\mu}$ to constrain the parameter space of the GMSB
models. We have presented bounds on supersymmetric particle masses as a function of $\tan \beta$. This combined
study has led to much stronger constraints on the parameter space of the model than those from either $b
\rightarrow s \gamma$ or $a_{\mu}$.  In particular, the region of large $\tan \beta$, which would have been
otherwise allowed, is ruled out or severely constrained, depending on the sign of $\mu$. With the inclusion of the
supersymmetric one-loop correction to $b$ quark mass, we have found that the region of large $\tan \beta$ and
negative $\mu$ is physically ruled out in order to give a correct value of $m_b$. The present experimental data on
the decay $b \rightarrow s \gamma$ leads to more stringent constraints than those from $a_{\mu}$ in a broad region
of the parameter space. The anticipated precision level of $4 \times 10^{-10}$ in determination of $a_{\mu}$ in
the Brookhaven E821 experiment would constrain the parameter space much more severely. \\

We thank Jim Smith and B. Dutta for helpful discussions.  This work was supported in part by the US Department of
Energy Grant No. DE FG03-95ER40894.

\newpage
\hspace*{5 cm} FIGURE CAPTIONS \\
\begin{itemize}
\item[Fig. 1~:] {Limits on the weak gaugino mass $M_2$ vs $|\mu|$ for $\tan \beta =10$, $\mu >0$ and $n=1$.
Units are in GeV.  The solid line represents the bound from the branching ratio for $b \rightarrow s\gamma$
(the region surrounded by the solid line is allowed) and the dot-dashed line represents the lower bound from $a_{\mu}$.
Note that $M_2$ and $|\mu|$ are calculated and not independent.  The value of $M_2$ increases as the value of
$|\mu|$ increases in the allowed region. }

\item[Fig. 2~:] {The same as Fig. 1, except $\mu <0$.}

\item[Fig. 3~:] {Limits on the weak gaugino mass $M_2$ vs $\mu$ for $\tan \beta =60$, $\mu >0$ and $n=1$.
Units are in GeV.  The solid line represents the bound from the branching ratio for $b \rightarrow s\gamma$
(the region surrounded by the solid line is allowed) and the dot-dashed line represents the lower bound from $a_{\mu}$.
Note that $M_2$ and $|\mu|$ are calculated and not independent.  The value of $M_2$ increases as the value of
$|\mu|$ increases in the allowed region.}

\item[Fig. 4~:] {The same as Fig. 1, except $n=3$.}

\item[Fig. 5~:] {The same as Fig. 2, except $n=3$.}

\item[Fig. 6~:] {The same as Fig. 3, except $n=3$.}

\item[Fig. 7~:] {Bounds on the sparticle masses (in GeV) as a function of $\tan \beta$ for $\mu >0$ and $n=1$.
The solid line represents the lower bound on the gluino mass, and the dotted and dot-dashed lines represent the
lower bounds on the stop ($m_{\tilde t_1}$ and $m_{\tilde t_2}$) and sbottom ($m_{\tilde b_1}$ and $m_{\tilde b_2}$)
masses, respectively. }

\item[Fig. 8~:] {The same as Fig. 7, except $n=3$.}
\end{itemize}

\begin{figure}[htb]
\vspace{1 cm}

\centerline{ \DESepsf(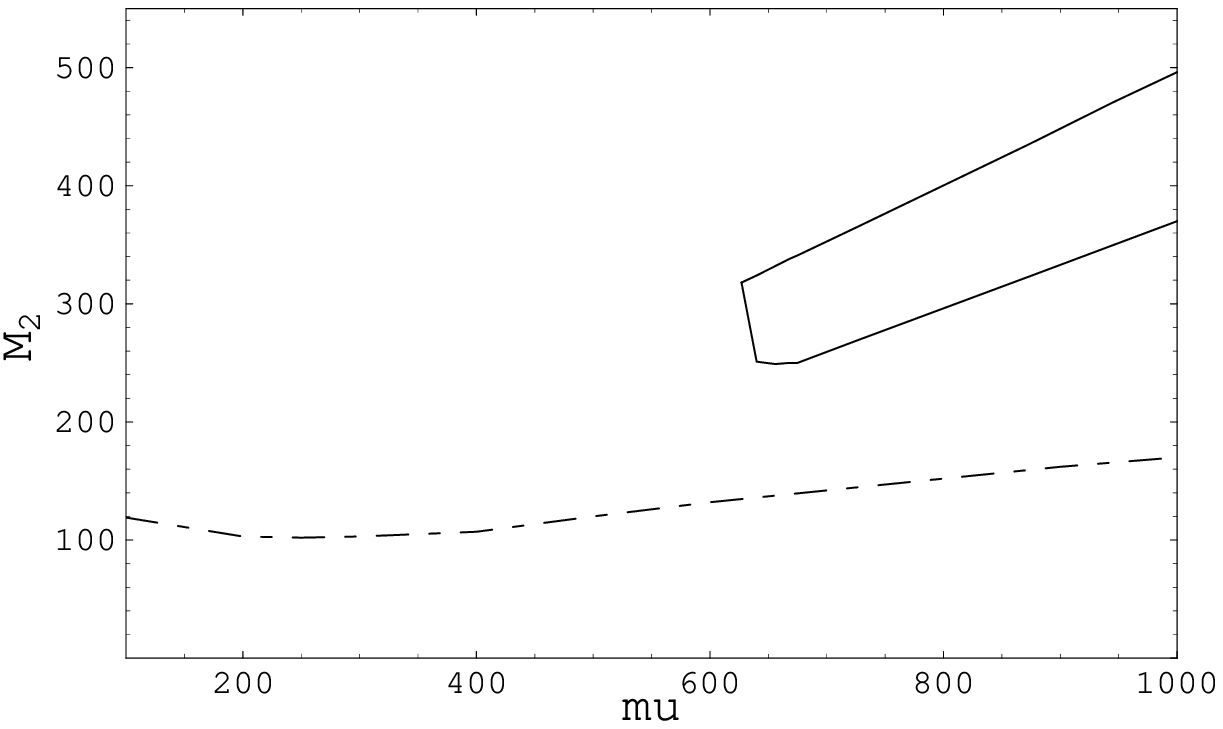 width 12 cm) }
\smallskip
\caption {} \vspace{3 cm}

\centerline{ \DESepsf(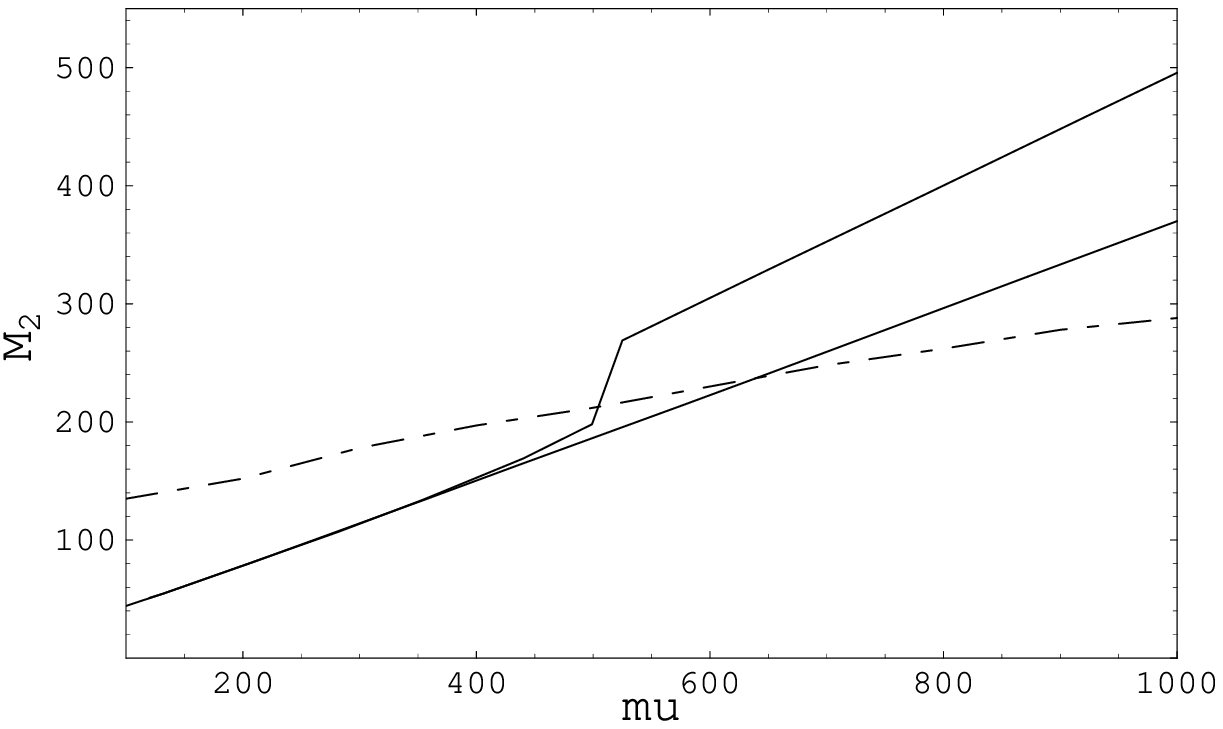 width 12 cm) }
\smallskip
\caption {} \vspace{3 cm}

\centerline{ \DESepsf(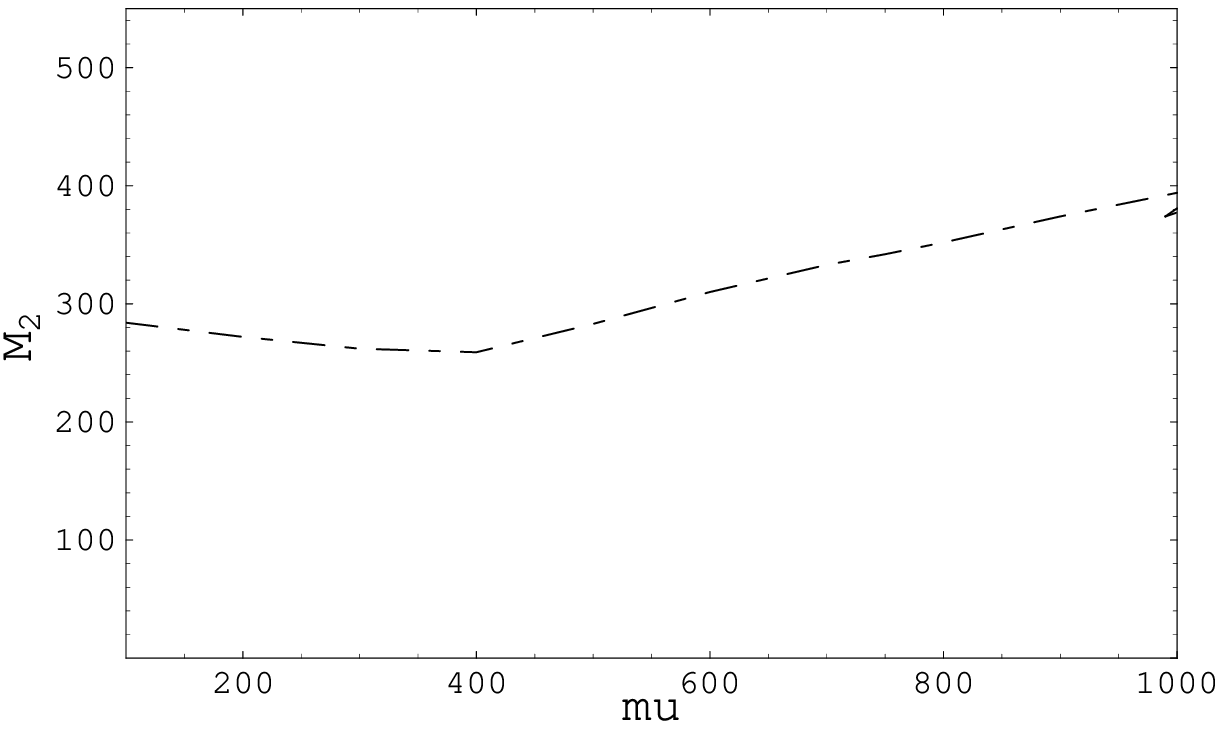 width 12 cm) }
\smallskip
\caption {} \vspace{3 cm}

\centerline{ \DESepsf(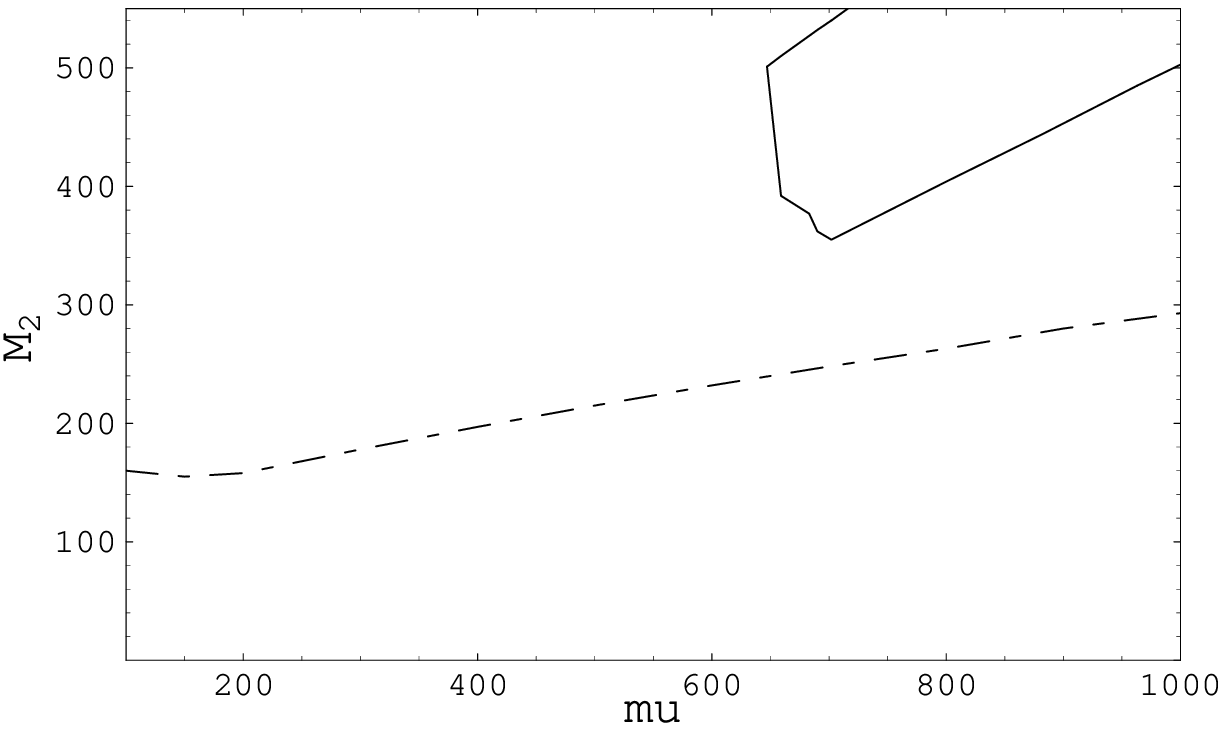 width 12 cm) }
\smallskip
\caption {} \vspace{3 cm}

\centerline{ \DESepsf(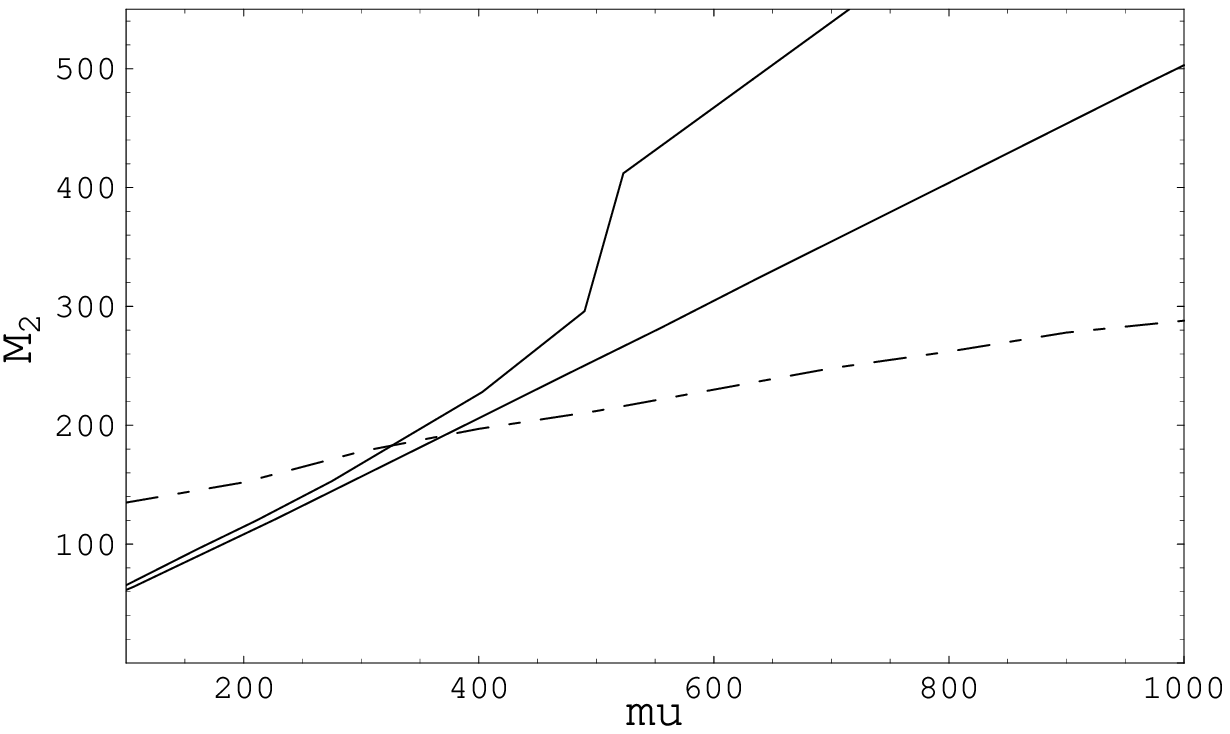 width 12 cm) }
\smallskip
\caption {} \vspace{3 cm}

\centerline{ \DESepsf(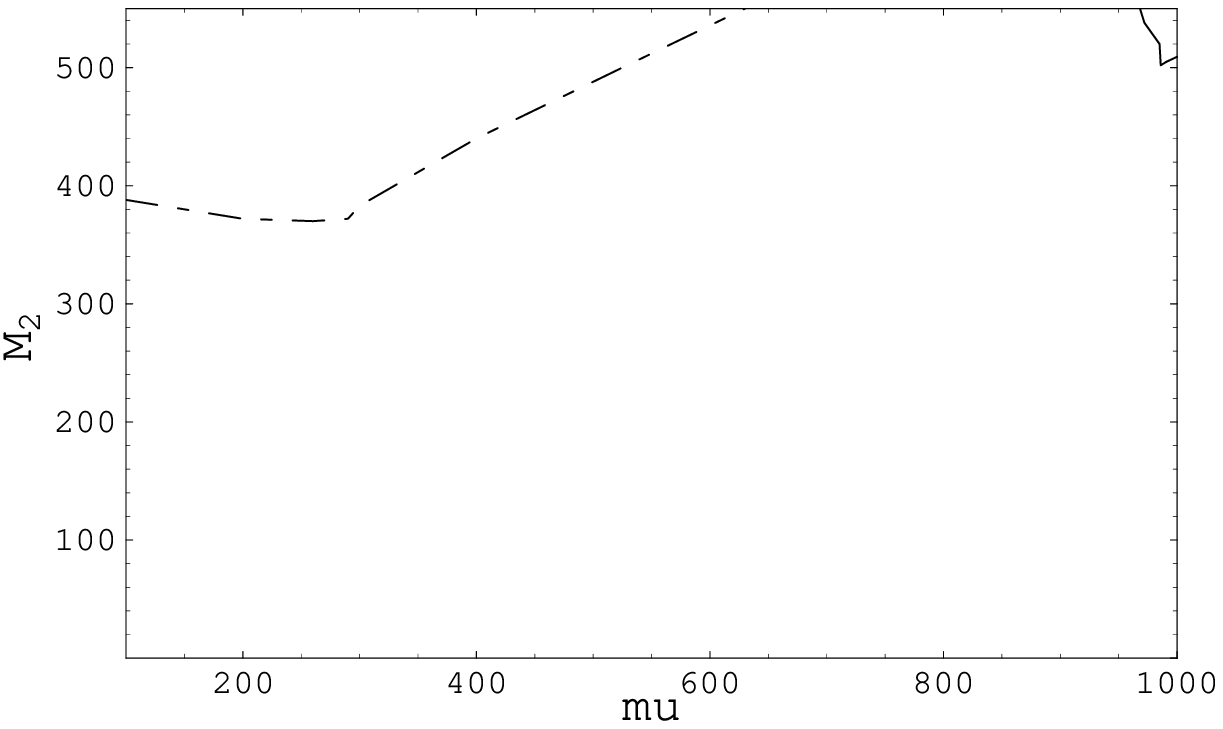 width 12 cm) }
\smallskip
\caption {} \vspace{3 cm}

\centerline{ \DESepsf(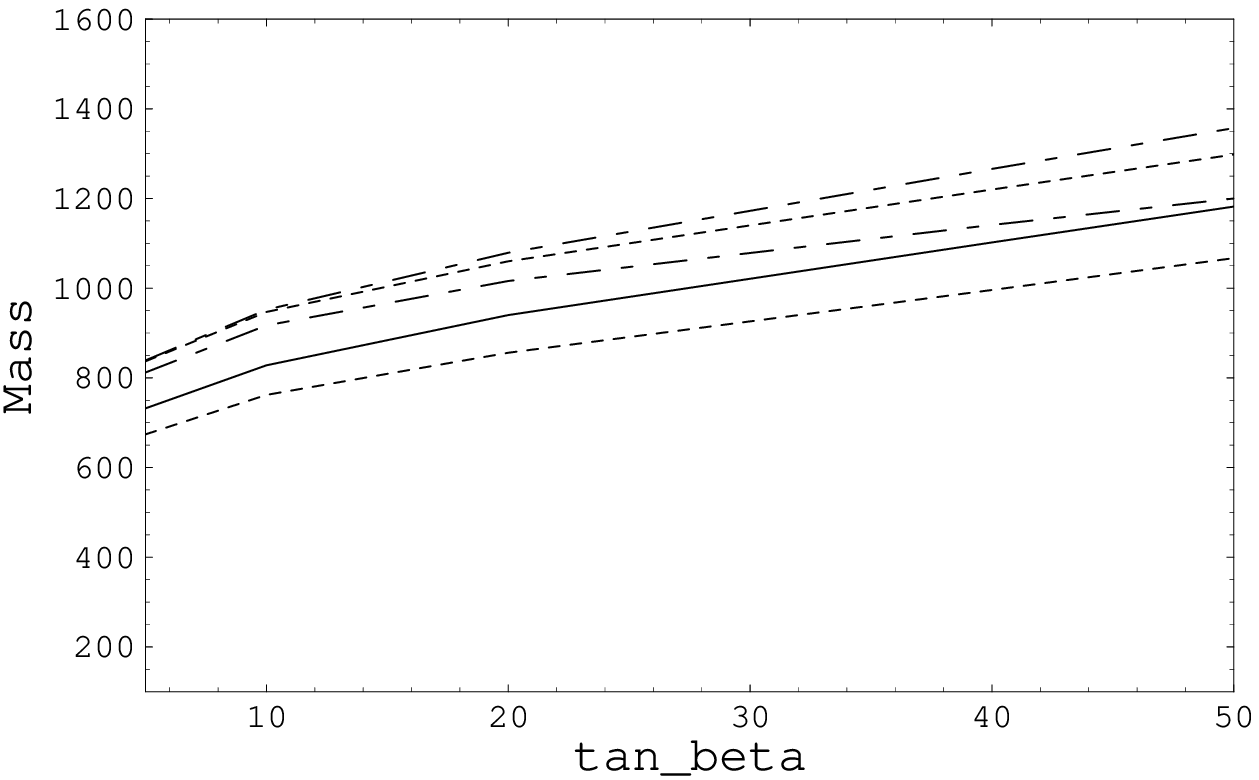 width 12 cm) }
\smallskip
\caption {} \vspace{3 cm}

\centerline{ \DESepsf(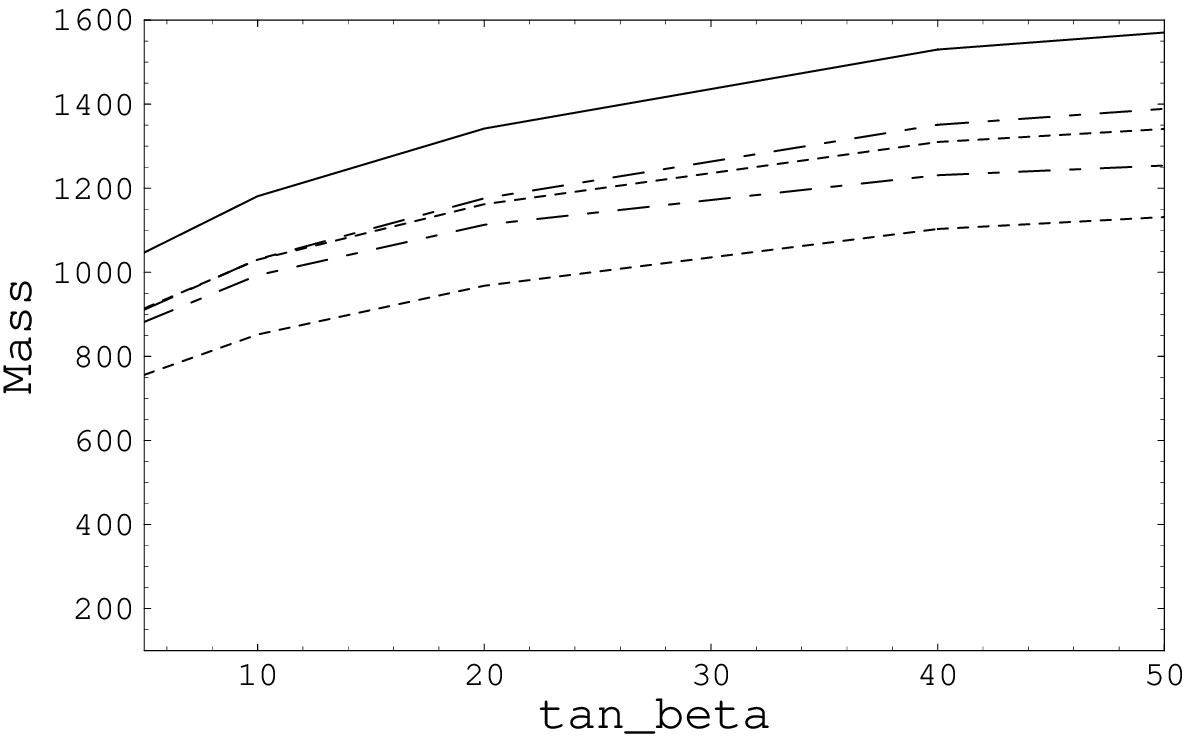 width 12 cm) }
\smallskip
\caption {}

\end{figure}
\end{document}